\documentclass[openacc]{rstransa}

\usepackage[acronym]{glossaries}
\usepackage{subcaption}
\usepackage{lmodern}
\usepackage{tcolorbox}
\usepackage{siunitx}
\usepackage{hyperref}

\usepackage{pgfplots,tikz}
\usetikzlibrary{positioning, calc, matrix}

\usepackage{newfloat} 
\DeclareFloatingEnvironment[
    fileext=lob,
    name=Box,
    placement=htbp,
    within=section
]{tcolorboxfloat}

\newacronym{awgn}{AWGN}{additive white Gaussian noise}
\newacronym{dgm}{DGM}{Deep Generative Model}
\newacronym{dps}{DPS}{Diffusion Posterior Sampling}
\newacronym{dm}{DM}{Diffusion Model}
\newacronym{ncsn}{NCSN}{Noise Conditional Score Network}
\newacronym{fmcw}{FMCW}{Frequency-Modulated Continuous Wave}
\newacronym{gan}{GAN}{Generative Adversarial Network}
\newacronym{nf}{NF}{Normalizing Flow}
\newacronym{vae}{VAE}{Variational Autoencoder}
\newacronym{ofdm}{OFDM}{Orthogonal Frequency-Division Multiplexing}
\newacronym{pmcw}{PMCW}{Phase-Modulated Continuous Wave}
\newacronym{map}{MAP}{Maximum a posteriori}
\newacronym{gas}{GAS}{Generative Adaptive Sampling}
\newacronym{cs}{CS}{Compressed Sensing}
\newacronym{sde}{SDE}{Stochastic Differential Equation}
\newacronym{gmm}{GMM}{Gaussian Mixture Model}
\newacronym{ddrm}{DDRM}{Denoising Diffusion Restoration Model}
\newacronym{mse}{MSE}{Mean Squared Error}
\newacronym{mmse}{MMSE}{Minimum Mean Squared Error}
\newacronym{ads}{ADS}{Active Diffusion Subsampling}
\newacronym{hdr}{HDR}{high dynamic range}
\newacronym{rf}{RF}{radio-frequency}
\newacronym{mri}{MRI}{Magnetic Resonance Imaging}
\newacronym{tr}{TR}{Repetition Time}
\newacronym{ista}{ISTA}{Iterative Shrinkage and Thresholding Algorithm}
\newacronym{nfe}{NFE}{Neural Function Evaluation}
\newacronym{mhz}{MHz}{Megahertz}
\newacronym{gbps}{Gbit/s}{Gigabits per second}
\newacronym{mbps}{Gbit/s}{Megabits per second}
\newacronym{vivit}{ViViT}{Video Vision Transformer}
\newacronym{convlstm}{ConvLSTM}{Convolutional LSTM}
\newacronym{adc}{ADC}{Analog-to-Digital Converter}
\newacronym{dl}{DL}{Deep Learning}

\newcommand{\F}{{\mathbf F}}
\newcommand{\w}{{\mathbf w}}
\newcommand{\x}{{\mathbf x}}
\newcommand{\y}{{\mathbf y}}
\newcommand{\n}{{\mathbf n}}
\newcommand{\z}{{\mathbf z}}
\newcommand{\A}{{\mathbf A}}

\newcommand{\I}{{\mathbf I}}
\newcommand{\R}{{\mathbb R}}

\newcommand{\boldeps}{{\boldsymbol\epsilon}}
\newcommand{\boldtheta}{{\boldsymbol\theta}}
\newcommand{\boldphi}{{\boldsymbol\phi}}
\newcommand{\norm}[1]{\left\lVert#1\right\rVert}
\DeclareMathOperator*{\argmax}{arg\,max}
\DeclareMathOperator*{\argmin}{arg\,min}
\DeclareMathOperator{\sign}{sgn}

\titlehead{Research}

\pgfplotsset{compat=1.18}

\begin{document}

\title{Deep Generative Models for Bayesian Inference on High-Rate Sensor Data: Applications in Automotive Radar and Medical Imaging}

\author{
Tristan S.W. Stevens$^{1}$, Jeroen Overdevest$^{1,2}$, Oisín Nolan$^{1}$, Wessel L. van Nierop$^{1}$, \\Ruud J.G. van Sloun$^{1}$, Yonina C. Eldar$^{3}$}

\address{$^{1}$Department of Electrical Engineering, University of Technology Eindhoven, the Netherlands\\
$^{2}$Signal Processing Algorithms, NXP Semiconductors, the Netherlands\\
$^{3}$Faculty of Mathematics and Computer Science, Weizmann institute of Science, Israel}

\subject{signal processing, electrical engineering, artificial intelligence}


\corres{Ruud J.G. van Sloun\\
\email{R.J.G.v.Sloun@tue.nl}}

\begin{abstract}
Deep generative models have been studied and developed primarily in the context of natural images and computer vision. This has spurred the development of (Bayesian) methods that use these generative models for inverse problems in image restoration, such as denoising, inpainting, and super-resolution. In recent years, generative modeling for Bayesian inference on sensory data has also gained traction. Nevertheless, the direct application of generative modeling techniques initially designed for natural images on raw sensory data is not straightforward, requiring solutions that deal with high dynamic range signals acquired from multiple sensors or arrays of sensors that interfere with each other, and that typically acquire data at a very high rate. Moreover, the exact physical data-generating process is often complex or unknown. As a consequence, approximate models are used, resulting in discrepancies between model predictions and the observations that are non-Gaussian, in turn complicating the Bayesian inverse problem. Finally, sensor data is often used in real-time processing or decision-making systems, imposing stringent requirements on, e.g., latency and throughput. In this paper, we will discuss some of these challenges and offer approaches to address them, all in the context of high-rate real-time sensing applications in automotive radar and medical imaging.
\end{abstract}


\begin{fmtext}

\end{fmtext}
\maketitle

\section{Introduction}
Active array sensing techniques are at the core of numerous advanced technologies, playing a critical role in fields such as automotive radar and medical imaging. These sensor arrays consist of multiple individual elements that actively emit signals and capture the reflected waves (sound or light) to obtain accurate representations of the scenery. The culmination of many sensory signals leads to massive data rates and corresponding challenges~\cite{zbontar2018fastmri, van2019deep, doris2022-radar-high-res}. These systems are required to provide high-resolution and real-time imaging, all while dealing with interference, noise and a changing environment.

Many of the challenges faced in array sensing can be effectively reformulated as inverse problems, which seek to estimate unknown parameters from observations. In the context of array sensing, this typically means reconstructing images or (distance, velocity or angular) measures from the raw data captured by the sensors. However, due to the underdetermined nature of these inverse problems, strong priors and knowledge of the underlying processes at hand are key for reliable reconstructions. This necessitates advanced modeling techniques that extract and capture meaningful information from the raw sensory data.

Recent advances in \acrfullpl{dgm} have shown great potential in solving problems with high-dimensional data by learning and exploiting the data manifold in domains ranging from medical imaging~\cite{chung2022score}, computer vision~\cite{rombach2022high} and natural language processing~\cite{dubey2024llama}. Specifically, they seek to model the distribution of data and subsequently sample from it, which can serve as a signal prior and aid in the inverse problem solving. However, applying these techniques directly to raw sensory data presents additional challenges due to the \acrfull{hdr} and rapid data acquisition rates, which impose stringent requirements on latency and throughput, further complicating the use of \acrshortpl{dgm}. Consequently, these challenges necessitate the development of tailored techniques involving generative models that effectively manage the unique characteristics of sensory data while adhering to the requirements of high-rate real-time sensing applications.

In this paper, we review ongoing work in application of deep generative models to sensory data. We start with some background on sensing applications, \acrshortpl{dgm} and finally Bayesian inference with \acrshortpl{dgm} in Section~\ref{sec:background}. Then, we discuss the two main challenges, namely model mismatches and real-time inference in Sections \ref{sec:model-mismatch} and \ref{sec:real-time}, respectively. In both of these sections, we discuss methods that mitigate these challenges, ranging from improved forward models with the use of \acrshortpl{dgm}, to accelerated inference techniques through approaches such as \emph{deep unfolding}, \emph{temporal inference} and \emph{active compressed sensing}.

\section{Background}
\label{sec:background}
\subsection{Sensing applications}
Many sensor modalities, such as medical imaging and radar sensing, share a common objective, namely the measurement of an \textit{unknown} channel impulse response. These channels are often deeply complex, with signals exhibiting significant dynamic range variations in combination with high sample rates. These characteristics of sensory data challenge accurate signal recovery. For instance, in both ultrasound imaging and radar, the generative model of the sensory data is not exactly known due to the highly variable reflectivity of objects within the sensor’s line of sight and the presence of multipath propagation. Extracting a signal-of-interest from raw data is a challenging task for existing model-based techniques, e.g., due to the lack of modeling capacity, non-Gaussian (structured) noise or other undesirable sensing phenomena such as interference or aberration. Another challenge faced when dealing with sensory data are high data rates, for example in real-time imaging applications. \acrfull{cs} \cite{candes2006near, eldar2012compressed, eldar2015sampling, rani2018systematic} has emerged as a powerful technique for reducing data rates by compressing the size of measurement vector $\y$ necessary to recover the signal-of-interest $\x$, through intelligent design of the sensing matrix $\A$ and exploitation of known signal statistics. A common application of \acrshort{cs} is via subsampling, which aims to recover the full signal-of-interest from a subset of possible measurements, typically fewer than required by the Nyquist-Shannon theorem. This has been successfully applied to reducing data rates in many domains, such as ultrasound imaging \cite{chernyakova2014fourier, ramkumar2019strategic, huijben2020learning}, radar \cite{nguyen2011robust, cohen2018sub}, and \acrfull{mri} \cite{ye2019compressed}.\\

\noindent Throughout this work, we refer to the following forward model:
\begin{align}
    \y  = \A\x + \n + \boldeps \ ,\qquad \qquad \x \in \R^M, \,\,\left\{\y, \n, \boldeps \right\}\in \R^N, \,\,\A\in\R^{N\times M}
\label{eq:forward-model}
\end{align}
where the observations $\y$ are contaminated with \emph{structured noise} $\n$ and thermal noise $\boldeps$, respectively. The thermal noise is assumed to be \acrshort{awgn}, i.e., $\boldeps \sim \mathcal{N}(\mathbf{0}, \sigma_{\epsilon}^2\I)$ where $\sigma_{\epsilon}^2$ represents its variance. In contrast, structured noise encapsulates all model errors (mismatch), including possibly non-linear effects, distortions and multipath components in the observations $\y$. The nature of $\n$ depends on the sensing scenario. In some applications, $\n$ may be independent of $\x$, such as structured interference from externals sources, commonly encountered in automotive applications of radar. However, in other scenarios, such as diffraction or multipath scattering in ultrasound imaging, $\n$ is inherently a function of $\x$, e.g. $\n(\x)$. Explicitly modeling (and performing inference on) the true forward physics model of $\n$ is often challenging. Given the complexity of capturing potential dependencies between $\n$ and $\x$, we assume independence, i.e., $p(\n|\x) \approx p(\n)$ and instead learn the marginal distribution $p(\n)$ in a fully data-driven fashion, as discussed in Section~\ref{sec:model-mismatch}.

For our goal of inferring the underlying signal-of-interest $\x$ from observations $\y$, we make use of the forward model in \eqref{eq:forward-model} combined with statistical priors. To establish these priors, we resort to \acrfull{dl} which has been proven to be effective for tasks that require accurate statistical models learned from the data itself. While black box approaches often fail when the trained networks are subjected to out-of-distribution data~\cite{shlezinger-model-based-dl}, recently, \acrfullpl{dgm} have shown exceptional capabilities when used in conjunction with Bayesian theory. By conditioning the generative process on observations, \acrshortpl{dgm} provide a robust framework for solving inverse problems, such as signal recovery in the presence of structured noise.

In the following, we provide a brief introduction to \acrshortpl{dgm} and their role in posterior sampling. We then highlight their growing use in various sensing applications, demonstrating their potential to enhance the accuracy and robustness of signal recovery in these challenging environments.

\subsection{Deep generative models}\label{sec:DGMs}
Generative models try to understand and model the underlying distribution of data and have an effective way of sampling new data points from this distribution. \acrshortpl{dgm} are a class of generative models that have specifically gained traction due to their ability to model high-dimensional data. At the core of \acrshortpl{dgm} are general parameterized function approximators, usually neural networks. These networks are trained on many examples from a training dataset, representing the data distribution. \acrshortpl{dgm} are able to effectively capture the structure of the data manifold as they leverage the property that all data points lie on a lower-dimensional manifold embedded in the high-dimensional data space~\cite{gorban2018blessing}.

Nonetheless, modeling distributions with \acrshortpl{dgm} poses several challenges. Probability density functions are constrained to be non-negative and integrate to one, which limits the choice of neural architectures. \acrfullpl{vae} \cite{kingma2013auto} circumvent the intractability of density estimation by approximating it with a variational lower bound. \acrfullpl{gan} \cite{goodfellow2014generative} are another class of \acrshortpl{dgm} that learn the data distribution implicitly through an adversarial objective. \acrfullpl{nf}~\cite{kingma2018glow} take a different approach altogether by transforming a simple base distribution into the target distribution through a series of invertible transformations.

Here, we focus on a more recent development in the field of \acrshortpl{dgm}, namely \acrfullpl{dm}. These models indirectly model the underlying distribution $p_\text{data}(\x)$ through the score function $\nabla_\x \log p_{\text{data}}(\x)$, which is the gradient of the probability density function with respect to the data itself. Unlike likelihood-based methods, this circumvents the need of directly modeling the probability density function. Furthermore, it leads to an interpretable denoising score-matching objective, which allows us to parameterize the score function with any neural network $s_\boldtheta(\x)$ and train it as follows:
\begin{equation}
    \argmin_\boldtheta{
        \mathbb{E}_{p_{\text{data}}(\x)} \left[ \norm{s_\boldtheta(\x) - \nabla_\x \log p_{\text{data}}(\x)}_2^2
    \right]}.
\label{eq:score_matching}
\end{equation}
This effectively results in a function that points back towards the data manifold and can be used to sample from the data distribution $p_{\text{data}}$. \acrshortpl{dm} model this sampling procedure through the reversal of a corruption process, also known as forward diffusion, which progressively adds increasing levels of \acrfull{awgn} until the sample is completely transformed from the original data distribution $\x_0\equiv\x\sim p_{\text{data}}$ to a Gaussian noise sample $\x_\mathcal{T}\sim\mathcal{N}(\mathbf{0}, \I)$, with diffusion time $\tau\in\left[0, \mathcal{T}\right]$. This continuous forward process $\x_0\rightarrow\x_\tau \rightarrow\x_\mathcal{T}$ can be formalized using a \acrfull{sde}:
\begin{equation}
\label{eq:forward-sde}
    \mathrm{d}\x = f(\tau)\x + g(\tau)\mathrm{d}\w,
\end{equation}
where $\mathbf{w}\in\R^d$ is a standard Wiener process, $f(\tau): [0, \mathcal{T}]\rightarrow\R$ and $g(\tau): [0, \mathcal{T}]\rightarrow\R$ are the drift and diffusion coefficients, which contribute to the deterministic and stochastic aspects of the \acrshort{sde}, respectively. Naturally, we are interested in the reversal of this process, which leads to the reverse diffusion process which has shown to result in a reverse-time \acrshort{sde} as follows \cite{song2020score}:
\begin{equation}
    \mathrm{d}\x =
    \big[
        f(\tau)\x - g(\tau)^2\underbrace{\nabla_{\x_\tau} \log{p(\x_\tau)}}_{\text{score}}
    \big] \mathrm{d}\tau + g(\tau) \mathrm{d}\Bar{\mathbf{w}}_\tau,
\label{eq:reverse_diff}
\end{equation}
where $\mathrm{d} \tau$ and $\mathrm{d} \bar{\w}$ are now processes running backwards in diffusion time. Conveniently, the \emph{score function} emerges from this reverse diffusion process and can accordingly be substituted with the learned score model from \eqref{eq:score_matching} to gradually remove noise and sample from $p_{\text{data}}$. Moreover, to facilitate the training process, the score model is conditioned on the diffusion time step $\tau$, resulting in a \acrfull{ncsn} $s_\boldtheta(\x_\tau, \tau)$ which is able to jointly evaluate the score of all perturbed data distributions $\forall\tau\in[0, \mathcal{T}]$ \cite{song2019generative}.

\subsection{Posterior sampling}\label{sec:posterior-sampling}
To reconstruct corrupted or incomplete incoming sensory data, according to \eqref{eq:forward-model}, using generative models, we resort to a probabilistic framework with deep generative models serving as foundation for inferring the underlying data. The act of posterior sampling centers around the idea of incorporating both prior information $p(\x)$ with incoming observations $\y$ according to Bayes' rule. Many posterior sampling algorithms have been proposed for various generative modeling architectures, for example the conditional Wasserstein \acrshort{gan} \cite{bendel2024regularized}. Here, however, we focus on posterior sampling with \acrshort{dm}s in order to provide the necessary background for the methods to follow.

As \acrshort{dm}s generate samples using gradients of probability density functions, we start by using Bayes' rule to formulate the posterior score function:
\begin{equation}
    \underbrace{\nabla_{\x_\tau} \log p(\x_\tau\mid\y)}_{\text{posterior}} = \underbrace{\nabla_{\x_\tau} \log p(\x_\tau)}_{\text{prior}} + \underbrace{\nabla_{\x_\tau} \log p(\y\mid\x_\tau)}_{\text{likelihood}}.
\label{eq:bayes}
\end{equation}
This expression factorizes the posterior distribution into a prior distribution which we model with \acrshortpl{dgm} and a likelihood term, which is a known distribution given our understanding of the physical acquisition process of the observed sensory data $\y$, capturing how the true signal $\x$ is corrupted by factors such as sensor noise, distortions and resolution limits.
In order to achieve posterior sampling with pre-trained \acrshortpl{dm}, one can substitute the score function in \eqref{eq:reverse_diff} with the factorization of \eqref{eq:bayes} leading to a conditional reverse-time diffusion process. The posterior score is then approximated as $\nabla_{\x_\tau} \log p(\x_\tau\vert\y)\approx s_\boldtheta(\x_\tau, \tau) + \nabla_{\x_\tau}\log{p(\y\vert\x_\tau})$.

Unfortunately, the structured noise-perturbed likelihood $p(\y\vert\x_\tau)$ is intractable, in contrast to the noiseless case $p(\y\vert\x_0)$. Various posterior sampling methods for \acrshortpl{dm} have been proposed to estimate this quantity \cite{song2023pseudoinverse, mardani2023variational, daras2024survey}. A widely used approach is \acrfull{dps}~\cite{chung2022diffusion}, which leverages the posterior mean that is derived via first order Tweedie's~\cite{efron2011tweedie}:
\begin{align}
    p(\x_0\vert\x_\tau) \approx \mathbb{E}[\x_0\vert\x_\tau] &= \frac{1}{\alpha_\tau}\left(\x_\tau + \sigma_\tau^2 \nabla_{\x_\tau} \log p(\x_\tau) \right) \\
    &\approx
    \frac{1}{\alpha_\tau}\left(\x_\tau + \sigma_\tau^2 s_\boldtheta \left(\x_\tau, \tau\right)\right):=\x_{0\mid\tau},
    \label{eq:tweedie_def}
\end{align}
where $\x_{0\mid\tau}$ represents the one-step denoising from diffusion step $\tau$. The first approximation corresponds to the \acrfull{mmse} estimator for $p(\x_0\vert\x_\tau)$~\cite{milanfar2024denoising}, while the second substitutes the score function with the trained \acrshort{ncsn}. Further, we reparameterize the \acrshort{sde} in \eqref{eq:forward-sde} using signal and noise rates, $\alpha_\tau$ and $\sigma_\tau$, which can be derived from the noise scheduling $f(\tau)$, $g(\tau)$ \cite{song2020score}, as $\x_\tau = \alpha_\tau \x_0 + \sigma_\tau \mathbf{\z}$ with $ \mathbf{\z}\sim\mathcal{N}(\mathbf{0}, \I)$. Finally, we approximate a tractable posterior score, by starting from \eqref{eq:bayes} and substituting the approximate gradient of the log likelihood using \eqref{eq:tweedie_def} as follows:
\begin{align}
    \nabla_{\x_\tau} \log p(\x_\tau\vert\y) &= \nabla_{\x_\tau} \log p(\x_\tau) + \nabla_{\x_\tau} \log p(\y\mid\x_\tau) \\
    &\approx s_\boldtheta(\x_\tau,
    \tau) + \nabla_{\x_\tau}\log p\left(\y\,\vert\, \x_{0 \mid \tau} \right).
\label{eq:approx-posterior}
\end{align}
The exact implementation of this posterior sampling framework varies based on the specific application. To illustrate this, we will provide examples from ultrasound, radar, and \acrshort{mri} in the following sections.

\section{Model mismatch}
\label{sec:model-mismatch}
\emph{Model mismatch} is a critical challenge in inverse problems involving real-world sensory data, where the assumed forward model deviates from the actual, often more complex, data acquisition process. While we often assume a known and accurate forward process $p(\y \mid \x)$ as described in \eqref{eq:forward-model}, this assumption rarely holds in practice.

\acrshortpl{dgm} have demonstrated strong performance in inverse problems when the forward model is fully known. However, in sensory data, the acquisition process often involves unknown propagation effects such as multi-path scattering, or sensor-specific distortions. Factors that are difficult to capture directly with a simple forward model or to learn directly from data using \acrshortpl{dgm}.

To close this gap, we discuss several key approaches. Firstly, in Section~\ref{sec:model-mismatch}\ref{sec:structured-noise}, we examine the concept of \emph{structured noise}, where model errors are explicitly captured with a \acrshort{dgm}, relaxing the reliance on a perfectly known forward model. Secondly, in Section~\ref{sec:model-mismatch}\ref{sec:hdr}, we address the challenge posed by the \acrfull{hdr} of raw sensory data, which can complicate the training and application of generative models. Finally, in Section~\ref{sec:model-mismatch}\ref{sec:model-based-score}, we incorporate \emph{model-based score functions} that leverage prior knowledge about the signal or sensing physics, enabling more robust guidance during inference. Throughout this section, the practical application of these concepts will be illustrated through two detailed examples in the domains of ultrasound imaging and automotive radar.

\subsection{Structured noise}
\label{sec:structured-noise}
One approach to dealing with model mismatch and other sensing and imaging artifacts is to model these error terms as structured noise. Logically, the structured noise cannot be captured using parametric probability distributions (such as Gaussian). Therefore, we resort to \acrshortpl{dgm} to learn its structure from data. This approach can effectively mitigate model mismatch in ultrasound and radar applications as we show in the following section.

For radar interference mitigation and multipath dehazing, similar source separation techniques have been applied through the use of joint posterior sampling using \acrshortpl{dgm}~\cite{overdevest-radar-interference-diffusion, stevens-ultrasound-dehazing}. The ill-posed problems are tackled by introducing two parallel generative processes that are conditioned on $\y$ to create a joint posterior sampling process using \acrshortpl{dm}~\cite{stevens2023removing}. Using Bayes' rule, samples are drawn from the joint posterior distribution $p(\x, \n \vert  \y)$ to obtain estimates of both signal and structured noise components, $\x$ and $\n$:
\begin{align}
    (\x, \n) &\sim p(\x, \n\vert \y) \propto p(\y\vert \x, \n) \cdot p(\x) \cdot p(\n) \ ,
\label{eq:joint-bayes}
\end{align}
where $p(\y\vert \x, \n)$ is the likelihood according to our measurement model in \eqref{eq:forward-model} and $p(\x)$ and $p(\n)$ are prior distributions which can be modeled using score-based \acrshortpl{dm} with the objective in \eqref{eq:score_matching}. The parallel posterior sampling (for both signal and structured noise components) is achieved by extension of \eqref{eq:reverse_diff}, through substitution of the score with the score of the joint posterior distribution. This results in the coupled diffusion process described by the following reverse-time \acrshort{sde}:
\vspace{-0.3cm}
\begin{equation}
    \mathrm{d}(\x_\tau, \n_\tau) = \bigg[ f(\tau)(\x_\tau, \n_\tau) - g(\tau)^2 \nabla_{\x_\tau, \n_\tau} \log p(\x_\tau, \n_\tau\vert \y) \bigg]\mathrm{d}t + g(t)\bar{\w}_t \ .
\label{eq:joint-reverse}
\end{equation}
We can again factorize the joint posterior using Bayes' rule for scores as follows:
\begin{align}
     \nabla_{\x_\tau} \log p(\x_\tau, \n_\tau \vert  \y) &= \nabla_{\x_\tau} \log p(\x_\tau) + \nabla_{\x_\tau} \log p(\y\vert \x_\tau, \n_\tau),\label{eq:posterior-x}\\
     \nabla_{\n_\tau} \log p(\x_\tau, \n_\tau \vert  \y) &= \nabla_{\n_\tau} \log p(\n_\tau) + \nabla_{\n_\tau} \log p(\y\vert \x_\tau, \n_\tau).
 \label{eq:posterior-n}
\end{align}
From this factorization it follows that each separate reverse diffusion process (for both signal and structured noise) is entangled through the shared joint likelihood term $\log p(\y\vert \x_\tau, \n_\tau)$. The prior score of the signal component $ \nabla_{\x_\tau} \log p(\x_\tau)$ can be either learned using a \acrshort{dm} or have an analytical prior such as sparsity. The structured noise score $\nabla_{\n_\tau} \log p(\n_\tau)$ is additionally learned using a separate \acrshort{dm} due to its complex nature. For a detailed description of the full algorithm we refer the reader to \cite{stevens2023removing}. In the following sections, we illustrate the practical application of these techniques addressing model mismatch, focusing on two major problems in the context of different sensing applications: ultrasound multipath scattering and radar interference.
\begin{figure}
    \centering
    \includegraphics[width=0.8\linewidth]{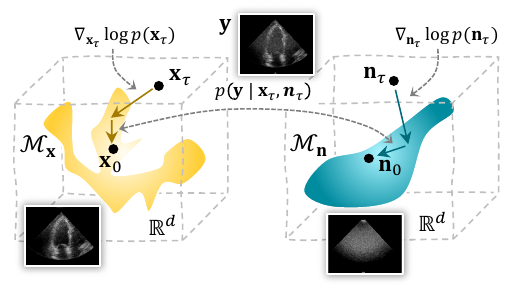}
    \caption{Overview of the proposed joint posterior sampling method for removing structured noise using \acrshortpl{dm}. During the sampling process, the solutions for both signal and structured noise move toward their respective data manifold $\mathcal{M}$ through the score functions. At the same time, the data consistency term derived from the joint likelihood $p(\y\vert \x_\tau, \n_\tau)$ ensures solutions that are in line with the (structured) noisy measurements. Figure adopted from \cite{stevens2023removing}.}
    \label{fig:enter-label}
    \vspace{-0.5cm}
\end{figure}

\subsection*{Example 1: Ultrasound multipath scattering}\label{subsec:ultrasound-mult}
Throughout this review, we will highlight several applications, demonstrating key techniques enabling the use of \acrshortpl{dgm} on sensory data. Each example is introduced with an information box, explicitly specifying the forward model and how \acrshort{dgm} is applied within the given context. Moreover, a short overview of data rates and sizes is provided to offer insight into the variability and characteristics of different types of sensory data.
\begin{tcolorboxfloat}
\refstepcounter{tcolorboxfloat}
\begin{tcolorbox}[title=Box~\thetcolorboxfloat\, -- Application: ultrasound imaging]
\textbf{Data characteristics:} A typical ultrasound probe comprises hundreds of individual transducer elements, each of which operates at sampling frequencies at the Nyquist rate, typically in the range of tens of \acrfull{mhz}, leading to $N$ fast-time samples per receive channel $C$. Depending on the transmit sequence (focused, diverging wave, plane wave) several hundreds slow-time sequences are acquired. In regular 2D brightness mode (B-Mode) imaging, at least several tens of frames per second can be expected, leading to raw data $\y\in \R^{C\times M\times N}$ rates that can quickly amount to several hundreds or thousands \acrfull{gbps}. This problem is exaggerated in 3D ultrasound, where matrix probes consist of thousands of elements~\cite{drori2021compressed}.\\
\textbf{Forward model:} $\y$ is the observed (and hazy) beamformed \acrshort{rf} data. $\x$ is the clean beamformed \acrshort{rf} data with the same dimensions, but only containing the direct path contributions. All clutter and multipath components are modeled through structured noise component $\n$ using a separate \acrshort{dgm}.\\
\textbf{Application of \acrshortpl{dgm}:} As both signal and haze contributions in the beamformed RF data are highly structured, \acrshortpl{dgm} can be fitted to both of these components.
\end{tcolorbox}
\label{box:ultrasound}
\end{tcolorboxfloat}

\begin{figure}
    \centering
    \includegraphics[width=1\linewidth]{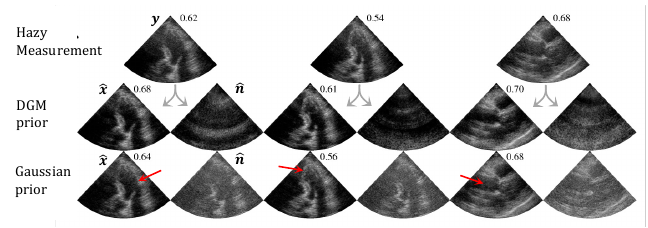}
    \caption{Comparison between a structured noise \acrshort{dgm} prior and a Gaussian prior for the task of dehazing \emph{in-vivo} medical ultrasound data. Posterior estimates of the signal $\hat{\x}$ and noise (haze) $\hat{n}$ are shown for each method, alongside corresponding gCNR \cite{rodriguez2019generalized} ($\uparrow$) values, highlighting the improved performance of the structured noise prior. Figure adopted from \cite{stevens2023removing}.}
\label{fig:comparison-dehazing}
\end{figure}

The first application we discuss is ultrasound imaging, a widely used modality in medical diagnostics due to its non-invasive and real-time nature. See Box~\ref{box:ultrasound} for a general overview of the application. Through the transmittance of high-frequency sound waves into the body, internal tissue structures can be reconstructed from the backscattered echoes. However, ultrasound signals are subject to a range of different noise sources that clutter the image and limit interpretability. One of the major origins for loss in image quality is caused by multipath scattering amidst layers of skin, fat and muscle between the transducer and the tissue being examined. These multipath reflections amount to a haze-like appearance on the image, dubbed simply \emph{haze}. Specifically, cardiac ultrasound is sensitive to haze due to the small transducer footprint and the addition of the ribs in line of sight of the probe.
In order to suppress the multipath clutter $\n$ and retrieve the direct path contribution $\x$ from the measured ultrasound signals $\y$ we consider the forward model in \eqref{eq:forward-model} and explicitly model both components separately with \acrshortpl{dm} \cite{stevens2023removing, stevens-ultrasound-dehazing}. For this purpose, we perform denoising score matching, see \eqref{eq:score_matching}, on (unpaired) training data samples of clean ultrasound $\{\x^{1}, \ldots, \x^{L}\}\sim p(\x)$, and multipath haze recordings $\{\n^{1}, \ldots, \n^{L}\}\sim p(\n)$ to learn two separate score functions, conditioned on the diffusion time step~$\tau$: \vspace{-0.2cm}
\begin{equation}
\nabla_{\x_\tau}\log p(\x_\tau) \approx s_\boldtheta(\x_\tau,\tau) \quad\text{and}\quad \nabla_{\n_\tau}\log p(\n_\tau) \approx s_\boldphi(\n_\tau,\tau).
\label{eq:dehazing-priors}
\end{equation}
Note that paired data of clean and hazy samples is not required to train these two generative models. This greatly reduces the difficulty of creating suitable datasets, as the structured noise can be acquired in isolation or simulated. See \cite{stevens-ultrasound-dehazing} for more details on the curation of the cardiac haze dataset. Moreover, learning each distribution with separate \acrshortpl{dgm} is more robust compared to a supervised method on paired data. The latter approach struggles with generalization due to the variability in paired samples and potential to overfit to specific instances of noise~\cite{stevens-ultrasound-dehazing}. During inference, the two trained \acrshortpl{dgm} can be deployed within the joint-posterior sampling framework as seen in \eqref{eq:posterior-x} and \eqref{eq:posterior-n}. The impact of a learned noise prior, compared to a traditional Gaussian prior on the problem of dehazing medical ultrasound data is illustrated in Fig.~\ref{fig:comparison-dehazing}. The learned prior yields improved contrast and clearer structural details, whereas the Gaussian prior leaves residual noise with structured components, suggesting it inadvertently suppresses parts of the underlying signal.

\subsection{High dynamic range}
\label{sec:hdr}
\begin{figure}
    \centering
    \begin{subfigure}[b]{0.4\textwidth}
        \hspace{-1cm}
        \includegraphics[height=5.8cm]{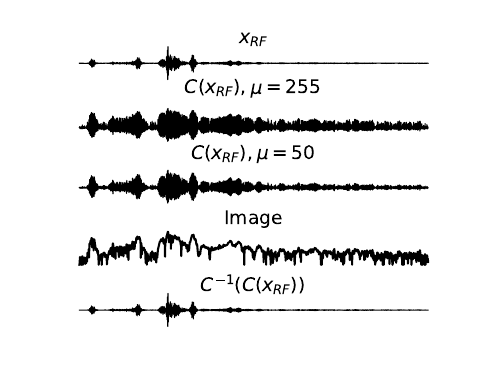}
        \label{fig:companding-signals}
    \end{subfigure}
    \hspace{2cm}
    \begin{subfigure}[b]{0.4\textwidth}
        \centering
        \includegraphics[height=5.5cm]{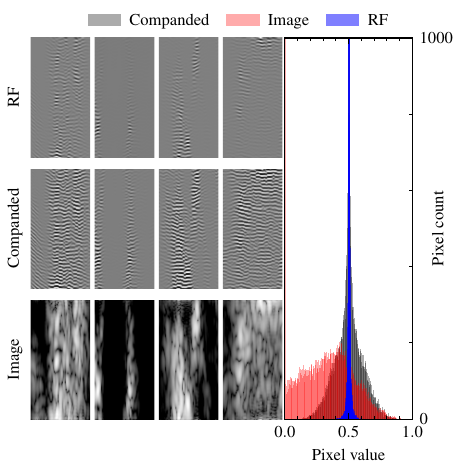}
        \label{fig:companding-histogram}
    \end{subfigure}
    \captionsetup{skip=-15pt}
    \caption{(Left) RF signals and their companded versions, where the $\mu$ value is adjusted to align the distribution with that of typical image pixel intensities. (Right) Histogram comparison of \acrshort{rf}, companded \acrshort{rf}, and ultrasound image data, illustrating the \acrshort{hdr} nature of \acrshort{rf} signals. Naturally, the companded \acrshort{rf} data exhibits a distribution more closely resembling that of image-domain data, facilitating more stable and effective training of \acrshortpl{dgm}. Figure taken from \cite{stevens-ultrasound-dehazing}.}
    \label{fig:companding}
\end{figure}

Unlike natural images, which typically have a relatively narrow range of pixel intensities, raw \acrfull{rf} data in sensor applications often exhibits a \acrfull{hdr}, meaning that the signal amplitudes can vary drastically, see Fig~\ref{fig:companding}. Besides this imposing constraints on the hardware side, requiring \acrshort{hdr} \acrfullpl{adc} \cite{azar2025unlimited}, this also presents challenges when training generative models such as \acrshortpl{dm}. The wide range of intensities can lead to numerical instability, with gradients either exploding or vanishing, and can cause the network to focus disproportionately on the stronger signals while neglecting weaker, yet important, components. In \cite{stevens-ultrasound-dehazing}, the \acrshort{hdr} of ultrasound signals is addressed through transformation of the \acrshort{rf} data using a technique known as \emph{companding}. This is an invertible operation that can \emph{compress} and \emph{expand} the dynamic range of a signal as follows:
\begin{align}
    \textit{compress:}\quad C(\x_{\text{RF}}) = \sign(\x_{\text{RF}})\frac{\ln({1 + \mu \vert \x_{\text{RF}} \vert)}}{\ln({1 + \mu)}}, \quad \textit{expand: }
    C^{-1}(\x) = \sign(\x)\frac{(1 + \mu)^{\vert \x \vert} - 1}{\mu},
\end{align}
with $-1\leq \x_{\text{RF}} \leq 1$, $-1\leq \x \leq 1$ and where $\mu$ is a parameter that determines the amount of compression applied. This ultimately leads to the following likelihood score:
\begin{align}
    \nabla_{\x_\tau, \n_\tau}\log \, p(\y\vert \x_\tau, \n_\tau) &\approx \nabla_{\x_\tau}\log \, p(\y \,\vert  \, \mathbb{E}[\x_0 \,\vert\, \x_\tau], \mathbb{E}[\n_0 \,\vert\, \n_\tau]) \label{eq:dehazing-tweedie}\\
    &= \zeta \nabla_{\x_\tau, \n_\tau} \norm{\y- C( \A\x_{\text{RF}, 0 \mid \tau} + \n_{\text{RF}, 0 \mid \tau})}_2^2 \nonumber\\
    &= \zeta \nabla_{\x_\tau, \n_\tau} \norm{\y - C(C^{-1}(\A\x_{0 \mid \tau}) + C^{-1}(\n_{0 \mid \tau}))}_2^2,\label{eq:dehazing-dc}
\end{align}
which can be used in combination with the two priors in \eqref{eq:dehazing-priors} modeled with diffusion networks $s_\boldtheta(\x_\tau, \tau)$ and $s_\boldphi(\n_\tau, \tau)$ to perform joint posterior sampling according to the framework in \eqref{eq:joint-reverse}. Note that we use Tweedie's formula from \eqref{eq:tweedie_def} to approximate \eqref{eq:dehazing-tweedie} and introduce $\zeta$ to group the constants as a result of the derivation of the data consistency term \eqref{eq:dehazing-dc}. These exact steps, including the companding technique, were used to generate the dehazed ultrasound images shown in Fig.~\ref{fig:comparison-dehazing}.

\subsection*{Example 2: Radar interference}\label{sec:model-mismatch-radar}
As a second case study, we consider the growing problem of \emph{mutual interference} in automotive radar, a domain where model mismatch arises from unpredictable signal interactions in increasingly congested environments. Mutual interference is becoming a major challenge in the automotive radar scene as more vehicles are being equipped with radar sensors, resulting in resource scarcity, i.e., time, frequency, and space~\cite{autom-radar-review}. Although the de-facto waveform currently implemented by the sensor manufacturers is \acrfull{fmcw}, whose chirp-like signals linearly increase or decrease its carrier frequency over time, there is currently no standardization present, leading to a free-for-all situation. A broad range of transmission schemes have been developed and implemented over time, i.e., up-down chirps, stepped-frequency, and frequency-coded waveforms~\cite{waveforms1,stepped}, resulting in a large variety of radar signals that can interfere with a victim radar. For more information see Box~\ref{box:radar}. This diversity burdens interference removal, causing simple existing mitigation strategies to become less effective. Model-based solutions~\cite{imat}, supervised neural networks~\cite{Mun20, Fuchs20, Ristea21, fuchs21}, and model-based deep learning methods~\cite{overdevest-radar-intf-deepunfold,overdevest-radar-intf-deepunfold2} have been proposed for interference removal. \acrshortpl{dgm} can learn a large variety of waveforms from training data to effectively remove interference signals; one implementation using posterior sampling is elaborated on below.

Generally, radar-to-radar interference mitigation occurs on the raw data directly, i.e., fast-time data, prior to any post-processing to avoid the interference from leaking into the other dimensions. Therefore, radar interference can be seen as structured noise, $\n$ in \eqref{eq:forward-model}, and leads to model mismatch with the interference being uncorrelated or correlated, ultimately reducing the radar's sensitivity or creating false positive detections. The authors in \cite{overdevest-radar-interference-diffusion} have used \acrshortpl{dgm} by applying score-based diffusion for solving \eqref{eq:posterior-x} and \eqref{eq:posterior-n} to separate the target reflections $\x$ and interference signals $\n$ from the observation $\y$.

\begin{tcolorboxfloat}
\refstepcounter{tcolorboxfloat}
\begin{tcolorbox}[title=Box~\thetcolorboxfloat\, -- Application: radar]
\textbf{Data characteristics:} Every radar is required, as defined in ISO standards, to send updates to the car of the objects present in the sensing environment (typically every \SIrange{40}{100}{\ms}, which includes the sensing time and the processing time of all downstream tasks). Therefore, real-time constraints are put on signal processing and deep learning solutions, such as interference mitigation, direction of arrival estimation, etc. Data rates in automotive radar can range from hundreds of \acrshort{mbps} to tens of \acrshort{gbps}, where a typical 3-D fast-time data cube $\y \in \R^{C\times M\times N}$ comprises $C$ receive channels, $M$ slow-time samples, and $N$ fast-time samples, respectively. With the recent trend towards high-resolution automotive radars, data cubes can for example grow towards $32 \times 256 \times 1024$, putting more stringent requirements on data rates, memory and computational load in the post-processing stages. \\
\textbf{Forward model:} $\y$ is the observed interfered signal, $\x$ is the complex-valued sparse signal containing the range information to all object reflections, and $\n$ and $\boldeps$ are the interference-only signal and thermal noise, respectively, in the same domain as $\y$. Therefore, we consider the following forward model: $\y=\F^\mathrm{H}\x + \n + \boldeps$, where we are interested in separating $\x$ and $\n$ from $\y$ using a \acrshort{dgm}.\\
\textbf{Application of \acrshortpl{dgm}}: To mitigate interference, we use a model-based and data-driven score-based \acrshort{dgm} to obtain estimates of $\x$ and $\n$, respectively. It is applied using only $N$ fast-time samples, for all $C$ channels and $M$ slow-time samples independently.
\end{tcolorbox}
\label{box:radar}
\end{tcolorboxfloat}

\subsection{Model-based score function}
\label{sec:model-based-score}
In contrast to Section~\ref{sec:model-mismatch}\ref{subsec:ultrasound-mult}, where two score functions are approximated using deep neural networks (see \eqref{eq:dehazing-priors}), specific signal properties, e.g. sparsity, can be inferred using model-based priors throughout the diffusion process to reduce complexity. As radar signals are known to be sparse in the range, Doppler and angular domain, the authors of \cite{overdevest-radar-interference-diffusion} exploit this domain knowledge in a model-based prior. Instead of having a deep neural network for approximating the score, $\nabla_{\x_\tau} \log p(\x_\tau) \approx s_{\boldtheta}(\x_\tau, \tau)$, as defined in \eqref{eq:dehazing-priors}, an analytical model-based score function is calculated
by rewriting \eqref{eq:tweedie_def} using Tweedie's approach. Then, the score function for the target signals at time step $\tau$  can be analytically defined as
\begin{equation}
   \nabla_{\x_\tau} \log p(\x_\tau) = \frac{1}{\sigma_\tau^2} \left(\alpha_\tau \x_{0\vert \tau} - \x_\tau \right) \ , \label{eq:score-targets}
\end{equation}
where $\x_{0\vert \tau}$ denotes the posterior mean, which the authors opt to obtain using the following $\ell_1$-norm minimization by promoting sparsity in $\x$:
\begin{align}
    \x_{0\vert \tau} :&= \arg\min_{\x} \frac{1}{2\sigma_\tau^2} \vert \vert \x_\tau - \x \vert \vert_2^2 + \lambda_\tau \vert \vert \x\vert \vert_1 \\ 
    &= \text{prox}_{\lambda_\tau||\cdot||_1}(\x_\tau) = \frac{\x_\tau}{|\x_\tau|} \left( |\x_\tau|-\lambda_\tau \right)_+ .\label{eq:soft-thr}
\end{align}
Formally speaking, this is known as a denoising step that is readily implemented using soft thresholding as shown in \eqref{eq:soft-thr} with $\lambda_\tau$ being time step-dependent. Additional benefits are that complex-valued score functions are avoided, which are generally hard to implement.

Furthermore, the authors have opted for a data-driven approach for approximating the interference score function $s_{\boldphi} (\n_\tau, \tau) \approx \nabla_{\n_\tau} \log p(\n_\tau)$, for which the network has been trained using the denoising score-matching objective of \eqref{eq:score_matching}. As explained earlier, the structured noise of the interference signals is challenging to analytically model due to its large waveform diversity, hence the use of conditional \acrshortpl{dgm} is favorable due to its generative ability. The distribution of the interference signals is learned in the raw data format directly.

\begin{figure}[h]
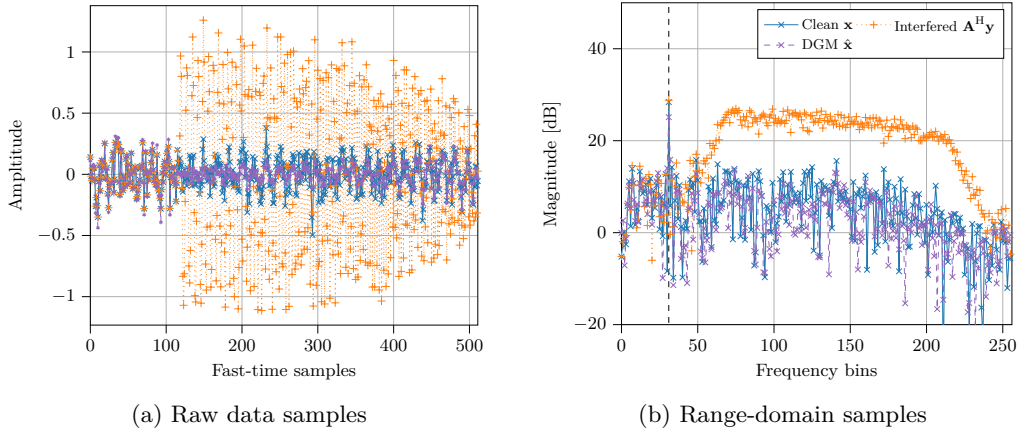

    \centering
    \begin{subfigure}[b]{0.48\textwidth}
        \centering
        \resizebox{\textwidth}{!}{\input{figures/interference/qual-res-td1}}
        \caption{Raw data samples}
        \label{fig:results-radar-intf-td}
    \end{subfigure}
    \hfill
    \begin{subfigure}[b]{0.48\textwidth}
        \centering
        \resizebox{\textwidth}{!}{\input{figures/interference/qual-res-fd1}}
        \caption{Range-domain samples}
        \label{fig:results-radar-intf-fd}
    \end{subfigure}
    \captionsetup{skip=-5pt}
    \caption{A semi-correlated interference scenario where a large portion of the raw data is contaminated, for which the capabilities of \acrshortpl{dgm} are shown. Figure taken from~\cite{overdevest-radar-interference-diffusion}.}
    \label{fig:results-radar}
    \vspace{-0.5cm}
\end{figure}

Under the guidance of the likelihood scores, using \acrshort{dps}, estimates of the targets and interference signals $\hat{\x}$ and $\hat{\n}$ are obtained using the joint posterior scores \eqref{eq:posterior-x} and \eqref{eq:posterior-n}, respectively:
\begin{align}
    \nabla_{\x_\tau, \n_\tau}\log \, p(\y\vert \x_\tau, \n_\tau) \approx \zeta \nabla_{\x_\tau, \n_\tau} \norm{\y - \F^\mathrm{H}\x_{0\mid\tau} - \n_{0\mid\tau}}_2^2.\label{eq:radar-dc}
\end{align}
In Fig.~\ref{fig:results-radar} the interference mitigation capabilities are shown for \acrshortpl{dgm} in a single target scenario for which a large part of the raw data is recovered from interference as shown in \ref{fig:results-radar-intf-td}, resulting in a reduced interference-induced noise-floor in Fig.~\ref{fig:results-radar-intf-fd}. Next, we explain how the methods of Examples 1 and 2 can be accelerated to enable the application of \acrshortpl{dgm} for real-time ultrasound probing and radar sensing.

\section{Real-time and high-data rates}
\label{sec:real-time}
Besides complex noise sources that cause model mismatches and impede image quality, excessive data rates and real-time conditions in sensor systems are another stringent requirement that incentivize efficient sensing and inference algorithms. Nevertheless, while extremely effective, \acrshortpl{dgm} are not generally known for their inference speed. In this section we discuss methods that either address this issue through development of accelerated methods that leverage \acrshortpl{dgm} in some way (Section~\ref{sec:real-time}\ref{sec:acceleration}), or try to reduce data rates through active compressed sensing (Section~\ref{sec:real-time}\ref{sec:acs}).

\subsection{Acceleration}
\label{sec:acceleration}
To bridge the gap between real-time inference of \acrshortpl{dgm} and high data-rates of sensing applications, our initial focus will be on acceleration of current methods. In order to maximize throughput in applications with high data-rates, efficient inference using \acrshortpl{dgm} is essential. Specifically, we will discuss temporal inference (Section~\ref{sec:real-time}\ref{sec:acceleration}-\ref{sec:temporal-inference}), deep unfolding (Section~\ref{sec:real-time}\ref{sec:acceleration}-\ref{sec:deep-unfolding}), and knowledge distillation (Section~\ref{sec:real-time}\ref{sec:acceleration}-\ref{sec:knowledge-distillation}).

\subsubsection{Temporal inference}
\label{sec:temporal-inference}
The sequential and real-time nature of sensory applications can be both a blessing and curse. While indeed the high throughput of data requires low-latency inference, the temporal axis can be exploited to accelerate inference and even improve reconstruction of the raw sensory signals. One way to accelerate posterior sampling methods using \acrshortpl{dm} as discussed in Section~\ref{sec:background}\ref{sec:posterior-sampling}, is to initialize a given diffusion trajectory conditioned on previous frames, effectively reducing the number of diffusion iterations necessary~\cite{stevens2024sequential}. Formally, given a set of $K$ diffusion posterior samples of previous frames $\x_0^{t-K:t}=\left\{\x_0^{t}, \x_0^{t-1}, ...,  \x_0^{t-K}\right\}$ we would like to estimate $p(\x^{t+1}\mid \x_0^{t-K:t})$ with some transition model (such as a \acrfull{convlstm} or \acrfull{vivit}) such that the number of diffusion steps necessary is minimized with $\tau^\prime\ll\mathcal{T}$. Rather than starting each diffusion trajectory from scratch at $\tau=\mathcal{T}$ with a Gaussian sample $\x_\mathcal{T}\sim\mathcal{N}(0, \sigma^2_\mathcal{T}\I)$, we use an appropriate estimate $\tilde{\x}$ based on past observations which we diffuse forward up to $\tau=\tau^\prime$ which leads to initialization of a shortened diffusion trajectory: $\x_{\tau^\prime}\sim \mathcal{N}(\alpha_{\tau^\prime}\tilde{\x}, \sigma^2_{\tau^\prime}\I)$. As shown in \cite{stevens2024sequential}, this reduces inference times for cardiac ultrasound imaging using \acrshortpl{dm} by $25\times$. Although in this case $\x$ represents B-mode images, applying sequential posterior sampling to raw sensory data could offer even greater advantages, which we consider an intriguing direction for future work.

\subsubsection{Deep unfolding}
\label{sec:deep-unfolding}
Deep unfolding (or deep unrolling) is a method that utilizes iterative model-based algorithms, such as proximal-gradient methods, in combination with neural networks to solve inverse problems. By unrolling the iterative optimization algorithm as a feed forward network it takes the structure of the iterations and allows for learning the parameters of the algorithm in successive steps. This will essentially apply multiple iterations in a single forward pass, thus accelerating the iterative algorithm at the cost of higher memory usage. Deep unfolding has been used in many real-world inverse problems such as sparse-coding~\cite{lista}, sub-Nyquist sampling~\cite{mulleti2023learning} and medical imaging \cite{li_deep_2021} but typically rely on discriminative networks. While there are prior works that combine deep unfolding with generative models \cite{wei-deep-unfolding, metz2017unrolledgenerativeadversarialnetworks}, there are none for high-data rate sensing applications, which could be a fruitful avenue to explore.

\subsubsection{Knowledge distillation}
\label{sec:knowledge-distillation}
Another powerful method to decrease inference time for efficient inference of \acrshortpl{dgm} is \textit{knowledge distillation}, in which a new \textit{student model} is trained to produce the same outputs as the original generative model using orders of magnitude fewer parameters. Knowledge distillation has been successfully applied to accelerate inference with both \acrshortpl{gan} \cite{aguinaldo2019compressing} and \acrshortpl{dm} \cite{salimans2022progressive}, among other architectures.

\begin{figure}
    \centering
    \includegraphics[width=1\linewidth]{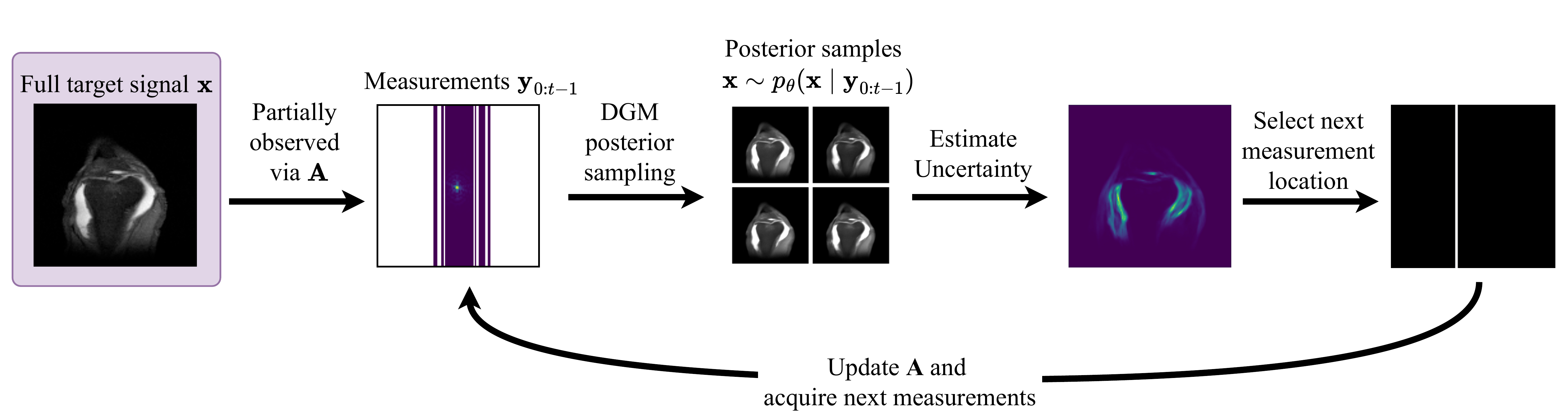}
    \caption{Illustrative example of an active compressed sensing step for \acrshort{mri} acceleration. The sensing matrix $\mathbf{A} = \mathbf{U}\F$ consists of a subsampling matrix $\mathbf{U}$ and DFT matrix $\F$. See \textit{fastMRI} \cite{zbontar2018fastmri} for more information about \acrshort{mri} acceleration.}
    \label{fig:active-compressed-sensing}
\end{figure}

\subsection{Active compressed sensing}
\label{sec:acs}
While \acrshortpl{dgm} have shown to produce excellent solutions to highly ill-posed \acrshort{cs} problems in many domains, e.g., medical imaging \cite{song2021solving, narnhofer2019inverse}, we focus here on \textit{active} \acrshort{cs} using generative models. Active \acrshort{cs} \cite{braun2015info} aims to sequentially design the sensing matrix $\A$ in real-time as the measurements are acquired, further compressing the required measurement vector and therefore of particular interest in applications with high data rates. Active \acrshort{cs} algorithms are considered `active' in the sense that they iteratively choose which measurements $\y_{t}$ to acquire next based on the measurements $\y_{0:t-1}$ they have observed so far. In the case of subsampling, for example, this involves choosing which measurement locations, e.g. pixels, time instances, antennas or other degrees-of-freedom, to sample in order to optimally reconstruct $\x$. Active \acrshort{cs} has a long history in various sensory applications, with existing works commonly using supervised deep learning \cite{van2021active} and reinforcement learning \cite{bakker2020experimental, stevens2022accelerated} to choose sampling locations. In the following, we review a number of recently proposed approaches to active compressed sensing using \acrshortpl{dgm} to jointly guide sampling and reconstruct target signals. In each case, the goal is to minimize uncertainty about $\x$, as measured by some uncertainty estimate derived from a set of posterior samples $p(\x \mid \y_{0:t-1})$ generated by the \acrshort{dgm}. A challenging aspect in this paradigm is estimating the decrease in uncertainty that will result from choosing a particular sensing design, leading to a variety of distributional and parametric assumptions in the methods discussed.

Sanchez \textit{et al.} \cite{sanchez2020uncertainty} propose \acrfull{gas}, a method for active subsampling in which the regions of the measurement signal with highest predicted variance are sampled next. This variance is computed by generating posterior samples of the full target signal $\x \sim p(\x \mid \y_{0:t-1})$ and passing them through the measurement model $\y_t = \mathbf{A} \x$ to generate samples of the full measurement signal at time $t$, yielding $\y_t \sim p(\y_t \mid \y_{0:t-1})$. The sample variance of this distribution over measurements $\mathbb{V}[\mathbf{A} \x \mid \y_{0:t-1}]$ can thus be computed directly from the posterior samples $\x$. The pixel or set of pixels maximizing this variance is then sampled at time $t$, and the algorithm repeats. Note that because the measurement noise is independent of the sampling location, choosing to sample the measurement location with the highest entropy will result in minimizing uncertainty about $\x$ \cite{van2023active}. Variance is however only proportional to the entropy under certain distributional assumptions, e.g. isotropic Gaussian. Despite this assumption, \acrshort{gas} proves to significantly outperform variable-density sampling in reconstructing MNIST images using a Wasserstein \acrshort{gan} \cite{gulrajani2017improved} as the \acrshort{dgm}.

Van de Camp \textit{et al.} \cite{van2023active} follow a similar approach to active subsampling, instead proposing the use of a \acrfull{gmm} to approximate the measurement posterior, i.e. $p(\mathbf{y}_t \mid \mathbf{y}_{0:t-1}) \approx \frac{1}{N_s}\sum_{i=1}^{N_s}\mathcal{N}(\mathbf{y}_t \mid \mathbf{y}_{0:t-1}, \I\sigma^2)$, where $N_s$ is the number of posterior samples. The measurement locations with maximum entropy are then selected to be sampled next, estimating the \acrshort{gmm} entropy via an approximation introduced by \cite{hershey2007approximating}. This entropy approximation is a function of the L2 distances between pixels across all pairs of posterior samples, leading to a regions with high 'disagreement' among samples being assigned high entropy. Van de Camp \textit{et al.} validate their sampling pipeline using two combinations of generative modeling architecture and posterior sampling technique to sample from $p(\mathbf{y}_t \mid \mathbf{y}_{0:t-1})$. In particular, they use (i) a \acrshort{vae} trained on MNIST \cite{lecun1998gradient} images, with Markov Chain Monte Carlo used to produce posterior samples in the latent space, which are decoded to generate full measurement samples $\mathbf{y}_t \mid \y_{0:t-1}$, and (ii) a \acrshort{gan} trained on \acrshort{mri} images from the \textit{fastMRI} \cite{zbontar2018fastmri} dataset, with annealed Langevin dynamics \cite{song2019generative} for posterior sampling.

Elata \textit{et al.} \cite{elata2024adaptive} propose \textit{AdaSense}, an adaptive compressed sensing method using a \acrfull{ddrm} \cite{kawar2022denoising} for posterior sampling. They propose using the \acrfull{mse} attained by the linear \acrfull{mmse} predictor as a measure of uncertainty. In the case where the values in $\mathbf{A}$ can be freely designed, \textit{AdaSense} uses the principle components of the posterior covariance as the rows of $\mathbf{A}$, leveraging that the principle components of the data covariance produce the linear \acrshort{mmse} predictor, thus minimizing uncertainty about $\mathbf{x}$. The algorithm proceeds iteratively by acquiring measurements $\mathbf{y}$ using $\mathbf{A}$, generating new posterior samples, and then adding the top $r$ principles components as new rows to $\mathbf{A}$, and repeating. Low values for $r$ then result in highly adaptive sampling, and vice versa for higher $r$ values. In many real world applications, however, $\mathbf{A}$ is constrained by the measurement process, and may not be freely designed. In these cases, AdaSense incorporates a new objective, aiming to minimize the linear \acrshort{mmse} when $\mathbf{A}$ is constrained to a set of possible sensing matrices $\mathcal{A}$, leading to the objective $\argmax_{\mathbf{A} \in \mathcal{A}} \mathbb{E}[(\mathbf{x} - \mathbb{E}[\mathbf{x} \mid \mathbf{y}_{0:t-1}])^\top \mathbf{A}^\dagger \mathbf{A}(\mathbf{x} - \mathbb{E}[\mathbf{x} \mid \mathbf{y}_{0:t-1}]) \mid \mathbf{y}_{0:t-1}]$. Here, $\mathbf{A}^\dagger$ is the Moore-Penrose pseudo-inverse of $\mathbf{A}$. For example, in \acrshort{mri} acceleration, the set $\mathcal{A}$ might consist of all possible next masks, where each next mask adds a new k-space line. \textit{AdaSense} is validated on \acrshort{mri} acceleration and natural image reconstruction tasks.

A limiting factor that is faced when using \acrshortpl{dm} to perform active sampling is the number of \acrfullpl{nfe} necessary to perform posterior sampling, due to the iterative nature of the reverse diffusion process. This may be in the range of hundreds to thousands for high quality image generation. Running an entire reverse diffusion process for each active sampling step may thus be infeasible for applications with high sampling rates. Nolan \textit{et al.} \cite{nolan2024active} address this problem with \acrfull{ads}, which performs $K < T$ active sampling steps in a single reverse diffusion process consisting of $T$ steps, resulting in a significant speedup for applications with large $K$. \acrshort{ads} uses \acrshort{dps} \cite{chung2022diffusion} as its posterior sampling engine, tracking an estimate of the posterior throughout the reverse diffusion process, and using it to select new measurement locations as it goes.
This estimate of the posterior is computed using a set of $N_s$ partially denoised samples $\{\x^{(i)}_\tau \mid \y_{0:t-1}\}_{i=0}^{N_s}$ at reverse diffusion step $\tau$, from which fully-denoised samples $\{\x_{0|\tau}^{(i)} \mid \y_{0:t-1}\}_{i=0}^{N_s}$ are computed via Tweedie's formula. \acrshort{ads} uses the \acrshort{gmm}-based approximation for $p(\y_t \mid \y_{0:t-1})$ proposed by Van de Camp \textit{et al.}~\cite{van2023active} to select maximum entropy sampling locations, validating the method on \acrshort{mri} acceleration as well as natural image subsampling.

\begin{tcolorboxfloat}[!h]
\refstepcounter{tcolorboxfloat}
\begin{tcolorbox}[title=Box~\thetcolorboxfloat\, -- Application: MRI]
\textbf{Data characteristics:} Of particular relevance regarding the data rate in \acrshort{mri} is the \acrfull{tr}, which measures the amount of time between successive pulse sequences on the same slice, and therefore determines the acquisition time for a single \acrshort{mri} slice. \acrshort{tr} tends to range from hundreds to thousands of milliseconds. In the popular \textit{fastMRI} \cite{zbontar2018fastmri}, for example, knee slices are acquired using \acrshort{tr}s in the range 2200-3000ms. \\
\textbf{Forward model:} A typical \acrshort{mri} scan images a 3D volume consisting of a set of stacked 2D \textit{slices}. \acrshort{mri} measurements are taken in the Fourier-space representation of the image, referred to as the \textit{k-space}, following the model $\y = \F\x + \mathbf{\epsilon}$, where $\mathbf{\epsilon}$ is measurement noise, $\x$ is the target image, and $\y$ are the k-space measurements. k-space measurements are typically acquired by a series of \textit{pulse sequences} for each slice, each of which provides a single line of points in the k-space. Given a full set of these k-space lines, the image can be recovered by the inverse Fourier transform. In accelerated \acrshort{mri}, the k-space lines are \textit{subsampled}, leading to the model ${\y_{0:t} = \textbf{U}\F\x+\mathbf{\epsilon}}$, where $\y_{0:t}$ are the measurements selected until time $t$ by subsampling matrix $\textbf{U} \in \mathbb{R}^{N \times M}$, i.e. $\A = \mathbf{U}\F$, with the goal of reconstructing the full image $\x$.\\
\textbf{Application of \acrshortpl{dgm}:} For \acrshort{mri} reconstruction \acrshortpl{dgm} are typically fit on fully-observed target images $\x$. The \acrshort{dgm} may, for example, be fit on complex-valued images $\x \in \mathbb{C}^{M}$ \cite{nolan2024active}, real images with a zeroed imaginary component \cite{elata2024adaptive}, or other variants \cite{van2021active}. Then, given such a \acrshort{dgm}, posterior sampling algorithms may be employed to recover full images $\x \sim p(\x \mid \y)$ from a subsampled k-space $\y$.
\end{tcolorbox}
\label{box:mri}
\end{tcolorboxfloat}

\subsection*{Example 3: Accelerated MRI}

\noindent As the final application, we consider accelerated \acrshort{mri}, a well-established use case for active \acrshort{cs}. For more context, see Box~\ref{box:mri}. Due to the relatively slow acquisition time (\acrshort{tr}) in \acrshort{mri}, inference time for popular \acrshort{dgm} architectures falls within real-time ranges, particularly when leveraging estimates of the posterior as in \acrshort{ads} or fast sampling algorithms such as \acrshort{ddrm} as in \textit{AdaSense}. The active \acrshort{cs} methods thus iteratively acquire k-space lines and update the subsampling matrix so as to select maximally informative next measurements, leading finally to a posterior distribution over fully reconstructed \acrshort{mri} images given the entire set of acquired measurements. This process is illustrated in Figure~\ref{fig:active-compressed-sensing}.

It remains an open challenge to accelerate such methods further to achieve real-time active \acrshort{cs} in domains with shorter acquisition times, such as ultrasound imaging, which may require a posterior inference time on the order of tens of milliseconds. However, with a combination of algorithmic and hardware advancements, this may soon be achievable.

\section{Conclusion}
Deep generative models are increasingly used to tackle problems involving high dimensional data. However, their integration with active array sensing applications poses unique challenges due to the need for real-time processing of the complex and dynamic nature of sensory data. Despite these hurdles, the potential gains of using of \acrshortpl{dgm} to enhance signal reconstruction by accurately modeling the underlying sensory data is significant. In this work, we highlight several works that aspire to close the gap between current \acrshort{dgm} capabilities and the demanding requirements of sensing applications, with the focus on two key areas: mitigating model mismatch through modeling of structured noise and addressing high-data rates and real-time processing through reduced inference times and active compressed sensing techniques. To this end, we showcase several illustrative applications that effectively apply \acrshortpl{dgm} for relevant problems in the domains of medical imaging and automotive radar. While the methods discussed in this review make substantial progress towards enabling real-time inference with \acrshortpl{dgm}, significant challenges remain in addressing the complexities of sensory data. Notably, applications involving extremely high data rates, such as 3D ultrasound and real-time radar systems, are not yet fully explored. To avoid a latency increase in the aforementioned real-time sensing applications, future work should optimize current methods by finding a balance between the powerful modeling capabilities of \acrshortpl{dgm} and the deployed acceleration techniques.

\bibliographystyle{RS} 
\bibliography{refs} 
\end{document}